\documentclass[pra,
               twocolumn,
                preprintnumbers,
                amsmath,
               amssymb,
                showpacs,
               superscriptaddress]{revtex4}
\usepackage{graphicx} 
\usepackage{dcolumn}  
\usepackage{bm}       
\usepackage{amsfonts}
\usepackage{dsfont}
\usepackage{xcolor}

\usepackage{ulem}
\usepackage{color}
 \newcommand{\Nadd}[1]{\textcolor{red}{#1}}

\begin{document}

\title{Direct Numerical Observation of Real-Space Recollision in High-Harmonic Generation from Solids}

\author{Mrudul M. S.}
\affiliation{%
Department of Physics, Indian Institute of Technology Bombay,
            Powai, Mumbai 400076, India }

\author{Adhip Pattanayak}
\affiliation{%
Department of Physics, Indian Institute of Technology Bombay,
            Powai, Mumbai 400076, India }

\author{Misha Ivanov}
\email[]{Mikhail.Ivanov@mbi-berlin.de}
\affiliation{%
Max-Born Institut, Max-Born Stra{\ss}e 2A, 12489 Berlin, Germany }
\affiliation{%
Blackett Laboratory, Imperial College London, London SW7 2AZ, United Kingdom}
\affiliation{%
Department of Physics, Humboldt University, Newtonstra{\ss}e 15, 12489 Berlin, Germany}

\author{Gopal Dixit}
\email[]{gdixit@phy.iitb.ac.in}
\affiliation{%
Department of Physics, Indian Institute of Technology Bombay,
            Powai, Mumbai 400076, India }

\date{\today}



\begin{abstract}
Real-space picture of electron recollision with the parent ion guides our understanding
of the highly nonlinear response of atoms and molecules to intense low-frequency laser fields.
It is also an important player in high harmonic
generation (HHG) in solids.  It is typically viewed in the momentum space,
as the recombination of the conduction band electron with the valence band hole,
competing with another HHG mechanism, the strong-field driven Bloch oscillations.
In this work, we use numerical simulations to directly test and confirm the real-space recollision picture as the
key mechanism of HHG in solids.  Our tests take advantage of the well-known
characteristic features in the molecular harmonic spectra, associated with the real-space structure
of the molecular ion. We show the emergence of analogous spectral features when similar real-space structures are
present in the periodic potential of the solid-state lattice. This work demonstrates the capability of HHG imaging of spatial
structures of a unit cell in solids.
\end{abstract}

\maketitle
\section{Introduction}
The importance of the simple physical picture
~\cite{corkum1993plasma, lewenstein1994theory, schafer1993above}
underlying high-harmonic generation (HHG) in atoms and molecules in strong laser fields can hardly be overstated. It plays a key role in using HHG to generate
and control attosecond pulses  ~\cite{paul2001observation, sansone2006isolated, popmintchev2012bright, goulielmakis2008single, kim2013photonic, li201753, silva2015spatiotemporal, gaumnitz2017streaking}.
It guides application of HHG as a time-resolved spectroscopic technique, linking the  harmonic
emission to the underlying electron-nuclear dynamics with attosecond temporal resolution
~\cite{krausz2009attosecond, smirnova2009high, dixit2012, bredtmann2014x, bucksbaum2007, corkum2007, lepine2014attosecond, baker2006probing, haessler2010attosecond, silva2013correlated, bruner2016multidimensional, smirnova2009attosecond, bian2010multichannel, morishita2008accurate}.
These dynamics  are
mapped on all properties of the emitted light: amplitude, phase, and polarization
~\cite{smirnova2009attosecond, raz2011vectorial}.

The advent of intense mid-infrared light sources
triggered experiments on high harmonic generation in solids, including dielectrics, semiconductors, nano-structures
and noble gas solids ~\cite{ghimire2011observation, ghimire2011redshift, zaks2012experimental, schubert2014sub,
vampa2015all, vampa2015linking, hohenleutner2015real, luu2015extreme, ndabashimiye2016solid, you2017high,
lanin2017mapping, sivis2017tailored, langer2018lightwave}.
HHG in solids is attractive both as a possible compact
source of XUV pulses~\cite{luu2015extreme,  vampa2017merge, kruchinin2018colloquium}
and as attosecond spectroscopy~\cite{silva2018high, silva2018all, bauer2018high, chacon2018observing, reimann2018subcycle, floss2018ab}. Spectroscopic applications include imaging energy
bands~\cite{vampa2015all,ndabashimiye2016solid}, performing tomography of impurities in solids~\cite{almalki2018high},
and unraveling electron--hole dynamics at their intrinsic timescale~\cite{higuchi2014strong,
ghimire2014strong, vampa2017merge, kruchinin2018colloquium}, including phase transitions, both light driven and
topological~\cite{silva2018high, silva2018all, bauer2018high, murakami2018high, chacon2018observing, reimann2018subcycle}.

The dominant
microscopic mechanism of HHG in solids is a topic of much debate, due in part to
very different (from atoms and molecules) dependence on laser and material parameters
~\cite{ghimire2011observation, ghimire2014strong, wu2015high, du2017quasi, tancogne2017impact, tancogne2017ellipticity, floss2018ab}.
One important mechanism is electron-hole recollision ~\cite{vampa2014theoretical},
leading to the (inter-band) electron-hole recombination
(see e.g. ~\cite{mcdonald2015interband,korbman2013quantum, vampa2015semiclassical, wu2015high, hawkins2015effect,
hohenleutner2015real}).
The second is the intraband current associated with laser driven Bloch oscillations (see e.g.
~\cite{hawkins2013role, ghimire2012generation, luu2015extreme, schubert2014sub}).

Here we describe results of the numerical experiment which allows us to  link directly HHG in solids with
real-space electron-hole recollision.
The key role of this mechanism, and the crucial importance of its real-space interpretation, has been highlighted in the beautiful recent paper~\cite{you2018probing},
we take advantage of  the
angstrom-scale spatial resolution embedded in the harmonic signal, well established in molecules~\cite{smirnova2009high, haessler2010attosecond, lein2002role, lein2002interference, lein2007molecular, vozzi2005controlling, kanai2005quantum, odvzak2009interference, torres2010revealing, sukiasyan2010exchange, yuan2014two}.
The spatial information arises from half-scattering during electron-molecule recombination.
It manifests in characteristic minima in the HHG spectra~\cite{lein2002role, lein2002interference, lein2007molecular}.
They are laser-intensity independent~\cite{smirnova2009high, lein2002role, lein2002interference, lein2007molecular} and are
associated with structure-based minima in the photorecombination cross
sections~\cite{lein2002role, lein2002interference, lein2007molecular, vozzi2005controlling, kanai2005quantum, odvzak2009interference, torres2010revealing, sukiasyan2010exchange, yuan2014two}, mirroring the well-known structure-related minima in photoionization.
In diatomic molecules, the minima result from the Cohen-Fano
interference of the two photoionization pathways originating at the two nuclei~\cite{cohen1966interference}.

Clearly, if real-space recollision between the conduction band
electron and the valence band hole underlies HHG in
solids, it is supposed to exhibit the same Cohen-Fano type interference
minima when the unit cell of the periodic lattice
potential has a two-centre structure.

\begin{figure}[h!]
\includegraphics[width=9 cm]{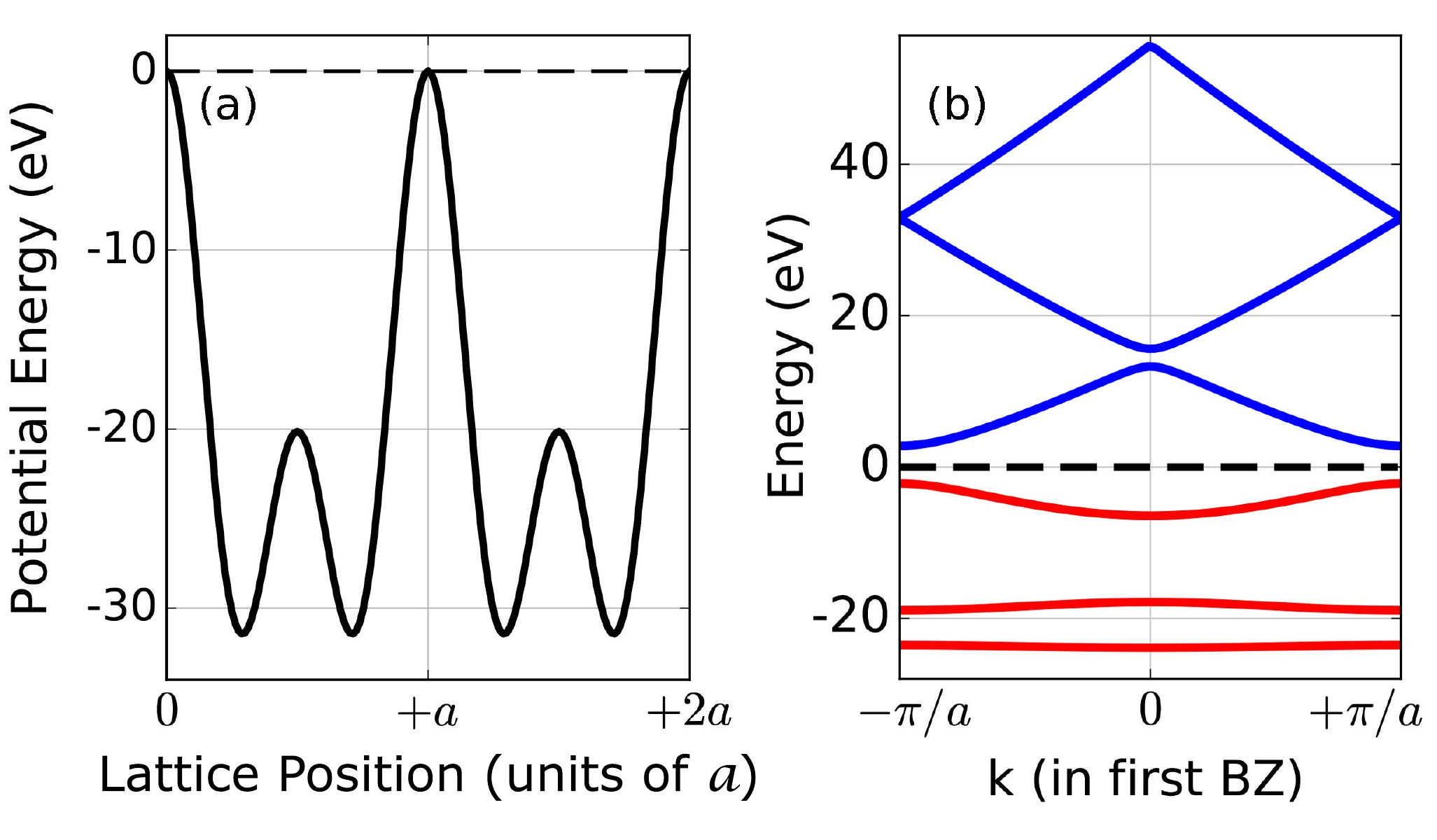}
\caption{ The periodic bichromatic lattice potential
[see Eq.~(\ref{eq02})] (a) and the corresponding energy-band structure with valence (red) and conduction (blue) bands marked (b)
within the first Brillouin zone.} \label{fig1}
\end{figure}

In dielectrics and wide-band-gap semiconductors, HHG is well
described using semiconductor Bloch equations describing effective single-particle motion in the
band-structure obtained, e.g., using density functional theory methods or  suitably chosen pseudo-potentials.
The quantitative accuracy of these methods has been well documented~\cite{hohenleutner2015real}.
In our study, we restrict ourselves to wide band gap materials, low-frequency drivers, and low excitation probability, where effective single-particle description is adequate.

\section{Theoretical Method}

To explore the idea, we model a semiconductor with two atom basis. In such a case, there will be two characteristic lengths for the unit-cell, the inter-atomic distance as well as the lattice constant. When the laser polarization is along the direction of the interatomic bond, we can effectively model this system using a one-dimensional bichromatic lattice potential (see Fig. 1(a)).
The harmonic spectrum was obtained by solving
the time-dependent Schroedinger equation (TDSE)
\begin{equation}\label{eq01}
i \frac{\partial}{\partial t} | \Phi(t)  \rangle = (\mathcal{H}_{0} + \mathcal{H}_{\textrm{int}}) | \Phi(t)  \rangle,
\end{equation}
where  $\mathcal{H}_{0} = -\nabla^{2}/2 + V(x)$ is the field-free Hamiltonian,  and
$V(x)$ is the bichromatic lattice potential
\begin{equation}\label{eq02}
V(x)  = -V_{0} \left[ (\alpha + \beta) - \alpha \cos\left(\frac{4 \pi x}{a}\right) - \beta \cos\left(\frac{2 \pi x}{a}\right)\right].
\end{equation}
Here, $V_{0}$ is the potential depth, $a$ is the lattice constant. Atomic units are used throughout unless stated otherwise.
Each unit cell  has a double-well shape, with $a$ the
distance between the unit cells (Fig. 1(a)), and the ratio of $\alpha$ and $\beta$ control the depth of the double-well potential.
In the present study,
$V_{0} = 0.37$ a.u., $a = 8$ a.u. and $\alpha = \beta = 1$ are used.   These values of  $V_{0}$
and $a$ are widely used to study HHG in solids~\cite{wu2015high, liu2017wavelength, liu2017time, ikemachi2017trajectory}.
Figure 1(b) shows the energy-band structure within the first Brillouin Zone for this bichromatic lattice.
The minimum energy band-gap is 4.99 eV at the edge of the Brillouin  zone ($k = \pm \pi/a$).

To obtain the high-harmonic spectrum,  we solve the TDSE in the velocity gauge within the Bloch-state
basis ~\cite{korbman2013quantum, wu2015high}, using the fourth-order Runge-Kutta method with 0.01 a.u. time-step
for the coupled differential equations as in Ref.~\cite{korbman2013quantum}, no ad-hoc
dephasing is introduced.

Initially, all the valence bands are expected to be filled.
Generally, one has to consider all crystal momentum states in the fully occupied valence
bands, adding coherently the resulting laser-induced dipoles. However, excitation
probabilities  are low, and we found that including only those crystal momentum states that
are located within    3$\%$ distance from the minimal bandgap, in the highest valence band
(third band in Fig. \ref{fig1} b), is sufficient to obtain the key spectral features of interest.
Other two valence bands which are deeply bound are neglected in the calculation. The spectra were converged for the number of conduction bands. Similar approximations are used in Ref.~\cite{wu2015high}.
To ensure the robustness of our findings, we have also simulated the HHG spectra  by
solving TDSE  in real space.
The results obtained from two different numerical approaches, in real space and in the Bloch state basis, show excellent agreement with
each other~\cite{future}.
The harmonic intensity is obtained from the time-derivative of the time-dependent current as
\begin{equation}\label{eq03}
I(\omega)  = \left|~\int dt~\left(\frac{d }{dt}j(t)\right)~e^{i \omega t}~\right|^{2}.
\end{equation}
Macroscopic propagation effects are not included.

\begin{figure}[h!]
\includegraphics[width=9 cm]{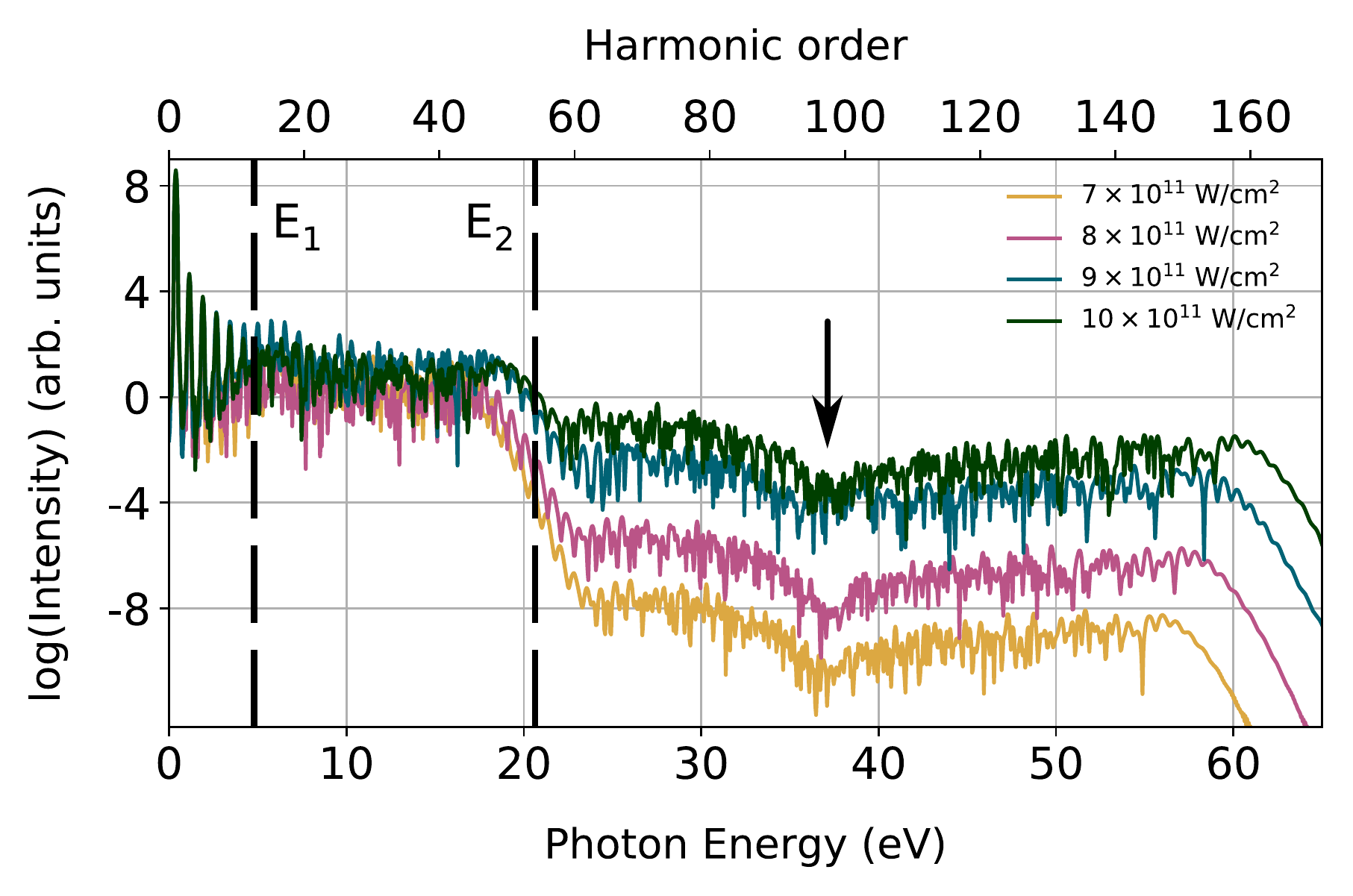}
\caption{ High-harmonic spectrum as a function
of driving laser intensity for the bichromatic lattice. Here,
$\textrm{E}_{1}$ and $\textrm{E}_{2}$  mark the minimum (4.99 eV)
and the maximum (19.67 eV) band-gaps between the first conduction band and the valence band. The black arrow represents the position of the minimum. \Nadd{Grey-scale values of plots are chosen in the ascending order of the laser intensity.}} \label{fig2}
\end{figure}

\section{Results and Discussion}

For the bichromatic lattice, the HHG spectrum is shown in Fig. 2, for an eight optical cycles linearly polarised laser pulse
with a sine-square envelope and $\lambda=3.2 \mu$m. Spectra corresponding to the
four different laser intensities are shown,
7 $\times$10$^{11}$ W/cm$^{2}$ (yellow), 8 $\times$10$^{11}$ W/cm$^{2}$ (pink),
9 $\times$10$^{11}$ W/cm$^{2}$ (blue) and 1 $\times$10$^{12}$ W/cm$^{2}$ (green).
The spectra  exhibit both a primary and a secondary plateau and
a sharp transition from the primary  to the secondary plateau, with clear cutoffs.
In the spectrum,
$\textrm{E}_{1}$ and $\textrm{E}_{2}$  mark the minimum
and the maximum band-gaps between the first conduction band and the valence band.

The primary plateau arises due to the
interband transition from the first conduction band  to the valence band.
The electron can also move to the higher conduction band
via interband tunnelling (see e.g. ~\cite{hawkins2015effect,ndabashimiye2016solid,ikemachi2017trajectory}.
Transitions from the higher-lying conduction band  to the
valence band   lead to the secondary plateau (e.g. ~\cite{ndabashimiye2016solid}).
The intensity of the second plateau increases with the laser
intensity, see Fig.  2, reflecting higher probability of the inter-band excitation to the higher conduction band.
The harmonic cutoff  in the second plateau increases linearly with the field amplitude of driving laser, as is
typical for solids ~\cite{ghimire2011observation, ghimire2014strong, wu2015high, du2017quasi}.

\begin{figure}[h]
\includegraphics[width= 9 cm]{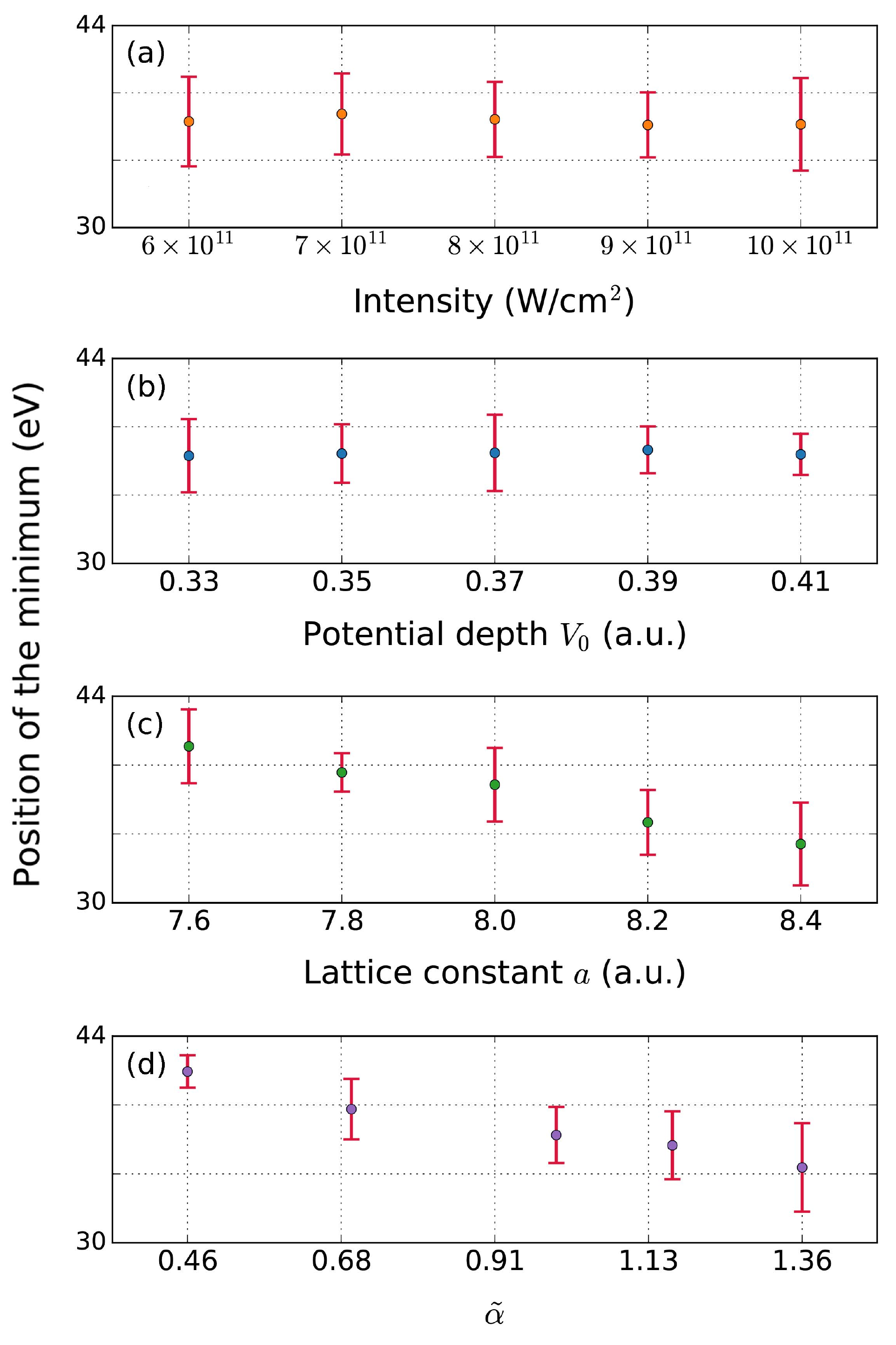}
\caption{The variation in the position of the minimum in the harmonic spectrum as a function of (a) intensity of the driving laser
($\lambda= 3.2 \mu$m, $V_{0} = 0.37$ a.u. and $a = 8$ a.u.),
(b) potential depth $V_{0}$ for the fixed lattice constant ($a = 8$ a.u.),
(c) as a function of the lattice constant ($a$) (for $V_{0} = 0.37$ a.u.); and (d) as
a function of the  depth of the double-well potential $\tilde{\alpha} = \alpha/\beta^{2}$
 (for $V_{0} = 0.37$ a.u., $a = 8$ a.u.).
In (b), (c) and (d),
the laser  intensity  $I=8 \times 10^{11}$  were used.
The error bar represents the  width of the  interference minimum.} \label{fig3}
\end{figure}

The key feature of interest is the pronounced minimum
in the second plateau, clearly present in Fig.  2 (see black arrow).
To identify its physical origin, we plot the
position of the minimum as a function of the laser intensity in Fig. 3(a).
It shows that the position of the minimum is not sensitive to the laser intensity, just like
the Cohen-Fano type interference minimum in high harmonic generation from molecules
~\cite{lein2002role, lein2002interference, lein2007molecular, kanai2005quantum, vozzi2005controlling, boutu2008coherent, zhou2008molecular, odvzak2009interference, torres2010revealing}.

To verify this conclusion, we look at the position of the minimum as a function of the
parameters of the bichromatic lattice potential [see Eq.~(\ref{eq02})].  As expected
for the Cohen-Fano type interference in radiative recombination during recollision,
the position of the minimum is independent of the depth of the bichromatic potential ($V_{0}$)
as long as the distance between the wells does not change, see
Fig. 3(b). However, the position of the minimum changes  as soon as we start to vary the lattice constant $a$, see Fig. 3(c).
As the lattice constant is increased, the minimum shifts  towards lower photon energies, as it should.
Identical observations have been reported for oriented
molecules, where the interference minimum occurs
at a lower harmonic order for larger internuclear bond-length,
(or when the aligned molecular ensemble is rotated towards the field polarization)
~\cite{lein2002role, lein2002interference}.
As can be seen from Eq. (2), the ratio of $\alpha$ and $\beta$
controls the depth of the double-well potential.
Changing this ratio changes the depth of the potential barrier between
the two wells, and thus the distance between the two peaks
of the corresponding wavefunction.
For small $\tilde{\alpha} = \alpha/\beta^{2}$,
the distance between
the two peaks in the double-well wavefunction is smaller, and so the
minimum is shifted to higher energies as evident from  Fig. 3(d).
Note that, the harmonic spectrum from single colour  lattice (monochromatic lattice) does not
exhibit any minimum in the spectrum (see also ~\cite{wu2015high, liu2017wavelength, liu2017time, ikemachi2017trajectory}).
Therefore, analogous to structural minimum in oriented molecules, this minimum in solid HHG is related to the structure of the potential. It is important to note that
the position of the minimum can be shifted to lower energies
by changing different parameters of the lattice potential.

\begin{figure}[]
\includegraphics[width=8 cm]{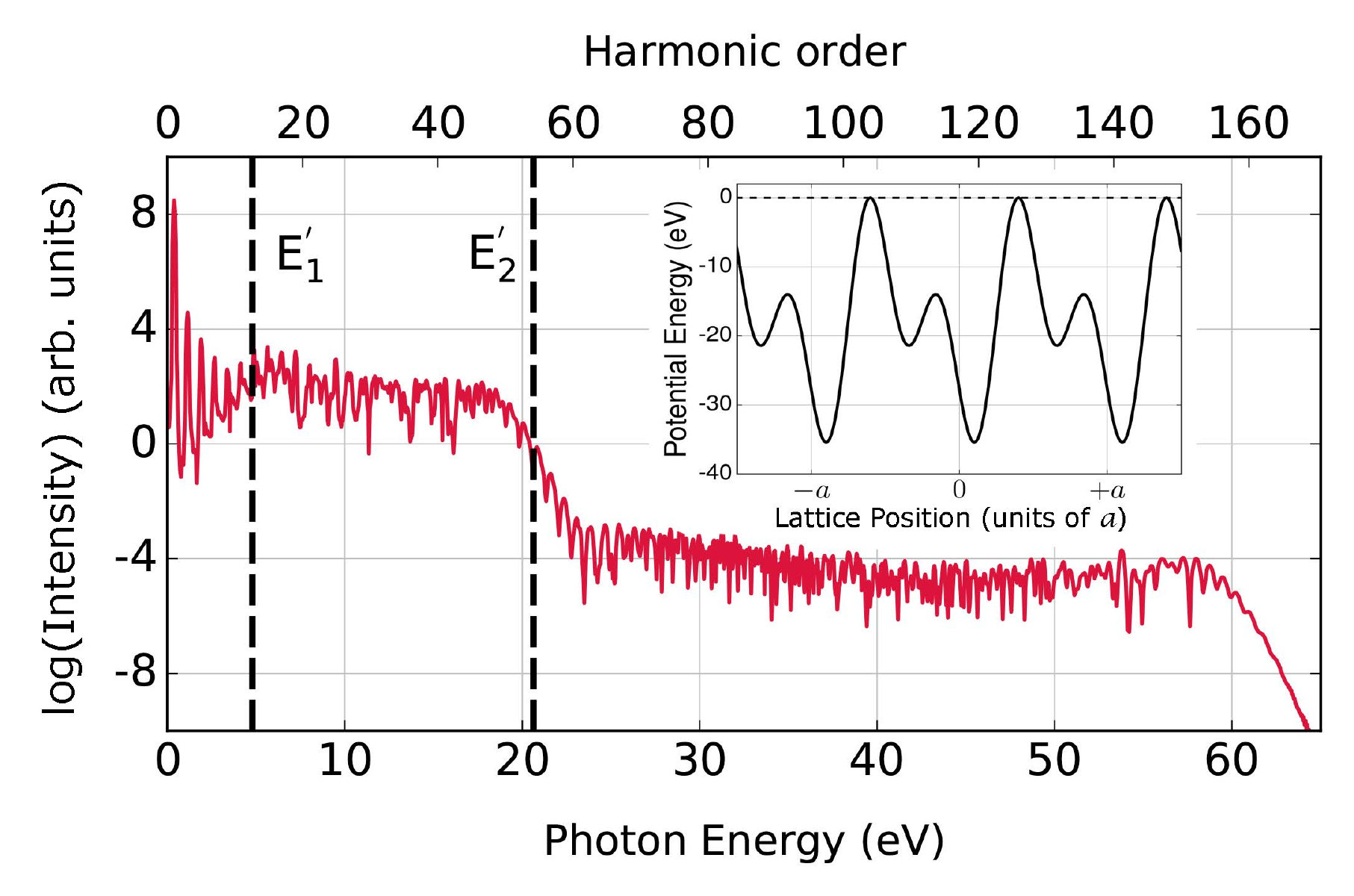}
\caption{High-harmonic spectrum for the asymmetic bichromatic periodic
lattice (shown in inset).  The harmonic spectrum is obtained for  intensity of  8 $\times$10$^{11}$ W/cm$^{2}$ and wavelength of  3.2 $\mu$m. Here,
$\textrm{E}^{\prime}_{1}$ and $\textrm{E}^{\prime}_{2}$  mark the minimum (4.79 eV)
and the maximum (20.63 eV) band-gaps between the first conduction band and the valence band.
The minimum energy band-gap is  at the edge of the Brillouin  zone ($k = \pm \pi/a$).} \label{fig4}
\end{figure}

In diatomic molecules,
the structural mimimum associated with photorecombination disappears when the two
nuclei are substantially different, so that the ground state is localized on a single nucleus.
The same should happen here.

To check this effect,  we introduce
asymmetry into the double-well potential of the lattice as shown in  Fig.~\ref{fig4} (inset).
The asymmetry is introduced by adding a 90$^{\circ}$ phase difference between the two spatial frequency
components of the lattice.
The corresponding harmonic spectrum is shown in Fig.~\ref{fig4} for I=8 $\times$10$^{11}$ W/cm$^{2}$ and
driving wavelength $\lambda=3.2 \mu$m. While the
overall harmonic spectrum is the same as for the symmetric bichromatic potential (see Figs. ~\ref{fig2} and ~\ref{fig4}), the
minimum disappears.
Therefore, the minimum in solid HHG does indeed represent the
structural minimum in recombination, in direct analogy with HHG in molecules, providing clear
evidence of the recollision picture of HHG in solids.

Let us further explore the underlying mechanism responsible for the presence of
minimum in a bichromatic lattice
and its absence in the monochromatic (single colour) lattice.
In molecules, where such structural minimum is well studied, the total potential of the molecule and the bound state wavefunction are written as a sum of
two components located at the two nuclei.
Consequently, the recombination amplitude also acquires two contributions,
associated with the
recombination onto each nucleus. Interference of these contributions leads to structural minima and maxima
in photo-ionization and in molecular HHG
spectra~\cite{lein2002role, lein2002interference, odvzak2009interference, torres2010revealing}. Note that while the exact position of the
structural minimum is sensitive to the details of the wavefunction,  the scattering potential, and possibly
multi-electron effects in photo-ionization or photo-recombination, the presence
of structure-induced features is completely general and is used to determine
molecular structures.

The same arguments apply in our case. While the monochromatic lattice potential
 is a  function of two reciprocal lattice points $(0, 2 \pi/a)$, the bichromatic potential is a function of three reciprocal lattice points
$(0, 2 \pi/a, 4 \pi/a)$ .
Let us write the dipole transition amplitude in the acceleration form as

\begin{eqnarray}\label{eq05}
\langle \Phi(t)| \nabla V | \Phi(t)\rangle  & = & -\frac{2 \pi V_{0}}{a} \left[
\langle \Phi(t)| \beta \sin \left(\frac{2 \pi x}{a} \right)  | \Phi(t)\rangle \right.  \nonumber \\
&& + \left. \langle \Phi(t)| 2 \alpha
\sin \left(\frac{4 \pi x}{a} \right) | \Phi(t)\rangle \right].
\end{eqnarray}

\begin{figure}[]
\includegraphics[width=8 cm]{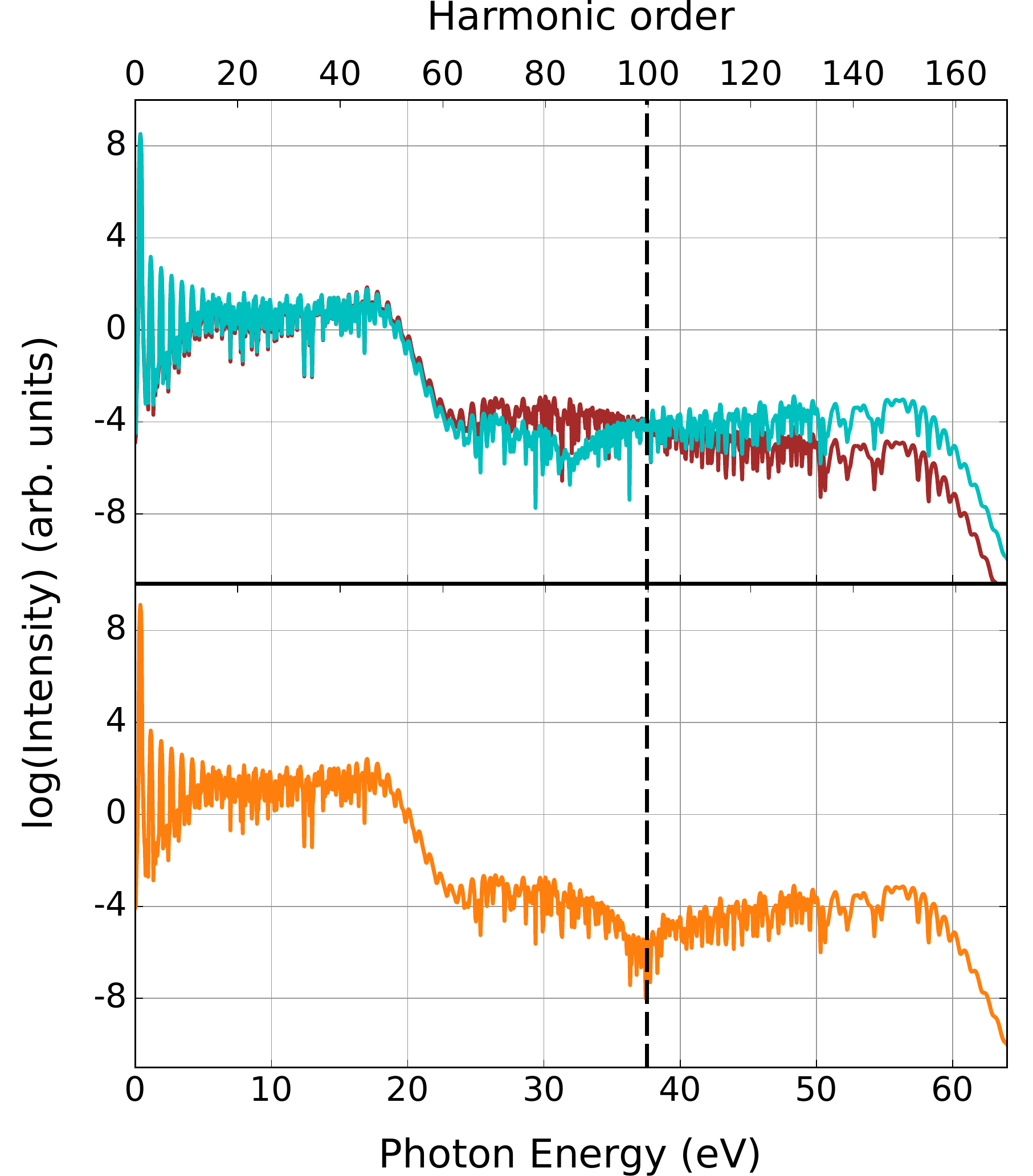}
\caption{Top panel:  High-harmonic spectrum corresponding to the two-components of Eq.~(\ref{eq05}): $-2 \pi V_{0}/a[\langle \Phi(t)| \beta \sin(2 \pi x/a) | \Phi(t)\rangle]$ (in brown  or with higher greyscale value) and
$-2 \pi V_{0}/a[\langle \Phi(t)| 2 \alpha \sin(4 \pi x/a) | \Phi(t)\rangle]$ (in cyan or with lower greyscale value).
Bottom Panel: The total High-harmonic spectrum corresponding to $\langle \Phi(t)| \nabla V | \Phi(t)\rangle $. The dotted line represents the region of the interference of the two-components and resultant minimum in the total spectrum. The harmonic spectrum is obtained for  intensity of  8 $\times$10$^{11}$ W/cm$^{2}$ and wavelength of  3.2 $\mu$m.} \label{fig5}
\end{figure}

The above equation shows that dipole transition amplitude is a linear superposition of the two-components,
which are functions of the two different reciprocal lattice points.
The interference of these two terms produces the structural minimum in the harmonic spectrum of solids. We
have calculated the harmonic spectrum from each of the
two components as shown in Fig.~\ref{fig5}. The first plateau of
the two structures are matching while the second plateau
behave differently for  the two components. The
point at which the two contributions match in the second plateau  can be identified
as the position of the structural minimum.
This also explains why there is no structural minimum in monochromatic lattice as there is no such two contributions to
interfere.
As it is clear from Eq.~(\ref{eq05}), only $a$ and  $\tilde{\alpha}$ are the parameters which make explicit changes in the two competing contributions differently, leading to a change in the position of the structural minima (See Fig.~(\ref{fig4})).
Also, Eq.~(\ref{eq05}) shows that the dependence of $V_{0}$ appears in the pre-factor.
While the components of $| \Phi(t)\rangle$ carry the information about $V_{0}$, this does not lead a considerable change in the position of the minimum for the
energy range that we have considered. By adding a phase difference in the two interfering terms in  Eq.~(\ref{eq05}), which is equivalent to creating an imbalance in the potential, we can modulate the interference as shown in Fig.~\ref{fig4}.

\section{Conclusion}
In this work, we have demonstrated that real-space recollision
picture passes an important numerical test, which makes a close analogy to the molecular picture. By providing direct numerical confirmation of
the key role of recollision in coordinate space, our work also suggests
that analysis of strong-field dynamics in solids
can benefit from real-space, as opposed to reciprocal space, perspective.
This might be particularly interesting when the real-space electron excursion
exceeds the size of the unit cell. Moreover, our work show that HHG from solid
has potential to image the internal structures of a unit cell in solids.

G.D. acknowledges the Ramanujan fellowship (SB/S2/ RJN-152/2015).
M.I. acknowledges support from the DFG QUTIF grant IV 152/6-1 and from the EPSRC/DSTL MURI grant EP/N018680/1.


\end{document}